# Detectors for Nuclear Physics

Tilak Kumar Ghosh [1,2]

[1] Variable Energy Cyclotron Centre , 1/AF, Bidhannagar, Kolkata -700064, India.

[2]Homi Bhabha National Institute, Training School Complex, Anushaktinagar, Mumbai-400094, India.

E-mail: tilak@vecc.gov.in

**Abstract:**
Progress in nuclear physics is driven by the experimental observation that requires state of the art detectors to measure various kinematic properties, such as energy, momentum, position etc. of the particles produced in a nuclear reaction. Advances in detector technology has enabled nuclear physicists to measure these quantities with better precision, and the reduced cost of the detection system has helped to have larger detection systems (array of detectors) to measure the rare processes with greater sensitivity. Several detection systems have been designed, developed and built in India over last few decades and are being used by the physicists. In this article, I will focus on such developments of detection systems at Variable Energy Cyclotron Centre (VECC), Kolkata.

**Introduction:**

It is a privilege to present this review on the detector developments for nuclear physics research at Variable Energy Cyclotron Centre (VECC), in this particular conference, that celebrates the centenary year of Bose Institute which happens to be the first inter disciplinary research institute in India, founded by Sir Jagadish Chandra Bose. He was one of the legendry Indian experimentalists who used to build his own instruments from the scratch for his research. The components of the instruments that he built (e.g; 60 GHz microwave apparatus [1] was quite innovative but remained pretty simple, that really attract and inspire us when we try to build our detectors in our laboratory.

**Scope of detection in fundamental nuclear physics:**

One of the fundamental questions that we deal in Nuclear Physics is, how do the nucleons work together to form complex nuclei. This branch of Nuclear Physics is often referred as *Nuclear Structure*. On the other hand, many of these nuclei are needed to make them react with target nuclei in order to understand the rearrangement of the nucleons inside nuclei that helps us to know the macroscopic or bulk properties of the nuclei. The interaction between two complex nuclei is broadly studied under the section called *Nuclear Reaction*. In general, the study of nuclear structure and nuclear reaction constitutes the low energy nuclear physics research. However there is another branch; the study of the constituents of nucleons, i.e. quarks and gluons and their interactions. This study requires very high energy to break the nucleons and this constitutes the high energy nuclear physics.

Let us have a look at the evolution of nuclear physics with energy (as shown in Fig 1). The change over from one zone to other is not sharp, rather having consid-erable overlap. However the classification is indeed helpful for designing experiments and relevant detector systems. The typical binding energy of a nucleon is ~ 8 MeV/u; so around that energy (below ~ 10 MeV/u) one can basically study the mean field dominated normal nuclear matter and the equilibrium reactions. Mean field gradually breaks down with increase in energy. Typically around the Fermi energy ( ~ 38 MeV/nucleon) one may expect transition from mean filed to nucleonic degrees of freedom which is indicated in the change of reaction mechanism. Low energy binary process are gradually replaced by multi-fragmentation process [2]. One may even expect exotic phenomena like nuclear liquid gas phase transition [3]. Beyond ~ 100 MeV/u, compressed nuclear matter may be produced where the main physics interest is to study the hydro dynamical flow and the dynamics of meson production. Beyond ~ 1 GeV/u it is possible to study the baryon production and at higher energy one can study physics of quark gluon plasma (QGP).

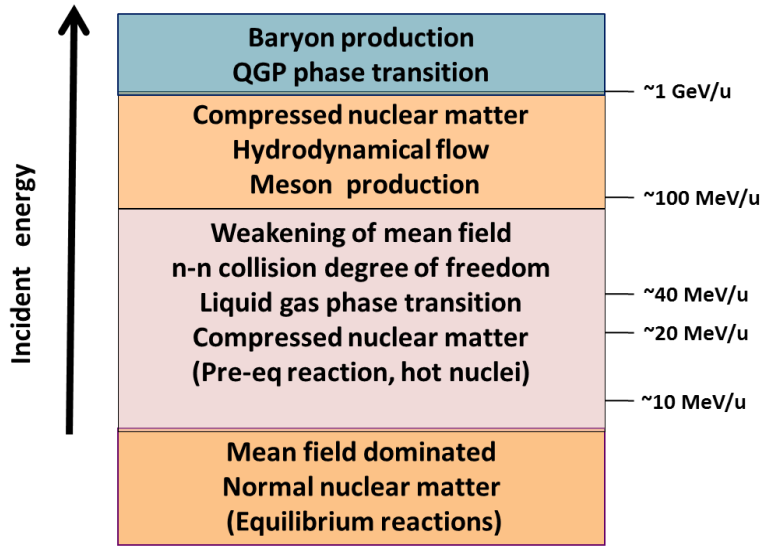

*Fig 1: Evolution of nuclear reaction scenario with incident energy*

With the increase in beam energy, the particle multiplicity, as well as the velocity of reaction products and variety of particles increase. Therefore from the detection point of view, at higher beam energy one would require detector system of large granularity and count rate handling capacity.

Usually four fundamental quantities are measured in an experiment and these are enough to extract the physics information about the nuclei and reaction process. These quantities are: energy ($E$), time of arrival ($t$), position ($x,y,z$) and multiplicity/counts (N) of a detected particle. For example, to know the atomic ($Z$) and mass number (A) of a particle produced in a nuclear reaction, measurement of particle energy loss ($dE/dx$) and time flight information are sufficient. So, a detec-tor with good time (*t*) and energy (*E*) resolution is required. The excitation energies of the residual nuclei and shapes of angular distributions can tell us about the reaction mechanism and properties of the residual nuclei. By measurement of energy (E) and angle (*x,y,z*) of the emitted fragments one can achieve that. By counting the number of particles (N) produced in a nuclear reaction, it is possible to calculate the cross section of the nuclear reaction that can inform us about a variety of properties of the nuclei as well as the reaction process.

In India, we have three major accelerator facilities: K130 Cyclotron at Variable Energy Cyclotron Centre in Kolkata, a 14UD pelletron machine at BARC-TIFR pelletron facility in Mumbai and a 15UD pelletron machine at Inter University Accelerator Centre (IUAC), New Delhi. Mumbai and Delhi machines are recently been upgraded with LINAC that provides beam up to Sulphur, typically less than 10 MeV/u. A K500 Super Conducting Cyclotron is under development at VECC.

At present different types of sophisticated detector systems are being developed at the various Indian accelerator centres. Since the list is extensive, it is not possible to describe all of them in this article. I shall concentrate on those detectors, which are more commonly used by experimental nuclear physics community and being developed by us. At our centre (VECC), the detector development activities are aimed at the effective utilization of the existing and upcoming accelerator facilities in the country.

**Development of gas detectors:**

Several types of gas detectors are being developed at VECC. One of the major physics research programs at our centre is the investigation of heavy ion induced fission reactions that has gained momentum in recent times. Apart from fundamental interest to study the dynamics of heavy ion nuclear reactions, these studies may guide us to find out the right kind of target and projectile combination for the synthesis of super heavy elements [4].

The detection of fission fragments is particularly suitable with gas detectors. Conventionally, segmented silicon detectors were used in past for detecting fission fragments but they have disadvantages like high costs, limited rate handling capacities, vulnerability to radiation damage and generally their smaller size. Multi wire proportional counter (MWPC), first developed by Georges Charpak [5], are more favorable in such experiments to detect fission fragments because of the flexibility it offers. Apart from being inexpensive, these detectors provide good timing and position resolution and are insensitive to radiation damage. Their ease of fabrication into any shape and size is an added advantage. They have good rate handling capacities and parameters like operating voltages and gas pressures can be fully customized to detect or reject light particles.

The detector we have developed is a low pressure MWPC. The active area of the detector is 20 cm x 6 cm. It consists of five wire planes: one anode (A), two sense wire planes (X,Y), two cathode (C) planes and in front of the detector there is an window frame which is basically a polypropylene foil (thickness ~ 0.5 micron). The wires of X and Y planes are of diameter 50 micron and fixed perpen-dicular to each other. The separation between two wires is kept 2 mm. The anode plane consists of wires with diameter 12.5 micron and separation between two wires is 1 mm whereas in case of the cathode planes the separation is kept same but the diameter of the wire used is 20 micron. The orientations of the cathode wires are also perpendicular to one another. Thin wires were taken, as smaller the diameter larger the field produced near the wire and hence the avalanches are localised. So the timing is faster and the dependence on the position of interaction in the detector becomes negligible. The gold coated tungsten wires are soldered on to the conducting strips. The two cathode wire planes are shorted outside and connected to a power supply through a charge sensitive preamplifier. This gives us the provision to get the energy loss signal from the cathode. In the X and Y planes, the wires are connected to the individual pads which are connected to successive pads of delay line chips. These delay line chips (each of 20 ns, 2 ns per tap) of the X position have a total delay of 200 ns with 10 chips. In case of Y position there is a total delay of 150 ns with 3 chips, each having a delay of 50 ns. The anode, cathode, X and Y are printed circuit broads (PCB) and spacers, made of a material called glass epoxy, are used to maintain the desired spacing of the detector. While the separation between anode and X (Y) sensitive planes are 1.6 mm, the separation between cathode and X(Y) sensitive planes are 3.2 mm. A 1 cm × 1 cm wire mesh of stainless steel wires of diameter 0.4 mm was used as a support to the polypropylene film. Two gas feed-throughs were connected to the back support frame which is made of stainless steel. Isobutane gas is continuously sent through the detector at a constant pressure flow mode with baratron feed-back closed loop flow control system (*make MKS, USA*). A schematic of the detector wire planes is shown in Fig 2.

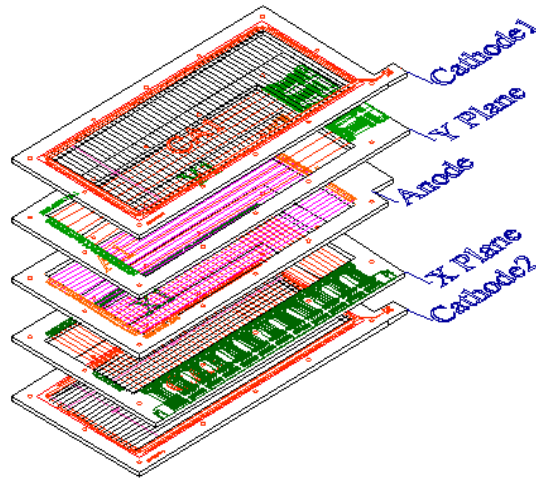

*Fig 2: Three dimensional view of the wire planes of a MWPC.*

In order to achieve the multi-step avalanche of electrons, the detectors are operated with a high electric field for substantial gas multiplication. Typical voltages applied on anode and cathode planes are +300 volts and -200 volts respectively. The operating gas pressure is about 2 torr. The reduced electric field $E/p$, where $E$ is the electric field between the cathode and sense wire plane, and $p$ is the gas pressure, are high enough to produce secondary multiplication of the primary electrons produced in the region between cathode and sense wires. However, the electric field in the cathode to anode region is a constant accelerating field. The constant filed in this region is not qualitatively changed by the introduction of the grounded X, Y sense wire planes at 1.6 mm distance. The intense field region around the central anode wire extends roughly twenty times the diameter, i.e., in this case about 0.25 mm.

The detectors were tested with a $^{252}$Cf source in the laboratory for uniformity of the position read outs and correspondence between the timing (anode pulse) and position (X-Y delay line signals). The position resolution (FWHM), measured by illuminating the detector with mask, was found to be 1.2 mm for X - direction and 1.5 mm for Y-direction. The anode signal for the fission fragments with a fast current sensitive preamplifier (ORTEC, VT120) is found to be 0.5 volts with rise time better than 5 ns. The typical time resolution of the detector, measured with $^{241}$Am source at a gas pressure of 5 torr is found to be 700 ps. The time resolution was measured by using one detector as a start and the 2$^{nd}$ one as a stop.

These detectors are routinely used for measurement of fission fragment mass distributions using the Indian accelerator facilities (K130 Cyclotron, BARC-TIFR and IUAC Pelletron) to study the fusion fission and quasi-fission dynamics [4]. Fig 3 shows a typical experimental set up where two gas detectors are placed at folding angle to detect the fission fragments. Pulsed beam from the accelerator are used as start signal and these detectors provide the stop signal in the time-of-flight setup to measure the mass distribution. The typical mass resolution achieved is 4 *amu*.

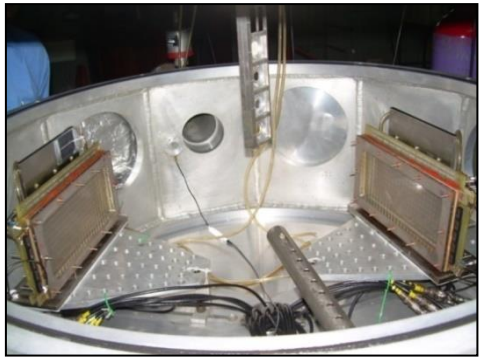

*Fig 3: Experimental setup inside VECC scattering chamber with the MWPCs to measure the fragment mass distributions in heavy ion induced fission reactions*

MWPCs of several dimensions were made in our laboratory as per the requirement of the experiment [6]. The MWPCs described above, can provide the mass of the detected fission fragments from measured time of flight information for binary reactions using set of equations *m1/m2= t1/t2* and *m1+m2 = M*, where *m, t* are the mass, time of flight of the binary fission fragments respectively and M is the compound nuclear mass. However, at higher beam energies typically in the Fermi energy domain, the reactions may not be binary in nature. Thus one would require both energy (*E*) and time of flight (velocity, *v*) information of the fragment in order to deduce the mass. This necessitates a detector that can simultaneously measure the energy and time of flight of the particle produced in a nuclear reaction.

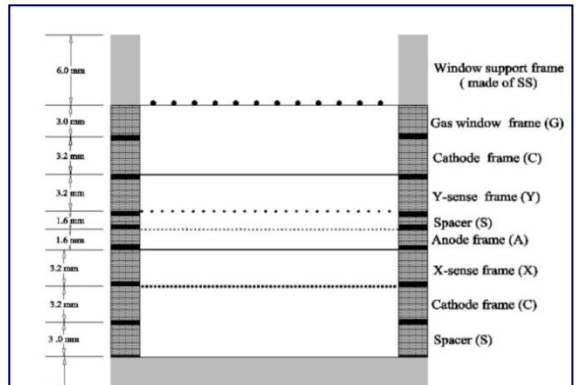

*Fig 4: Vertical cross sectional view of the hybrid gas detectors*

As a spinoff of MWPC development, we have developed a hybrid gas detector for simultaneous measurement of energy and time. The detector is conceptually a combination of low pressure MWPC and an assembly of segmented silicon detector. The cross sectional view of the detector is shown in Fig 4. The active area of the detector is 20 cm x 6 cm. A 300 micron thick segmented silicon detector assembled at the back of the MWPC has 16 strip each with dimension 30 mm x 96.8 mm. While the timing and position information are achieved from MWPC, the total energy of the particle is taken from the silicon detector
.

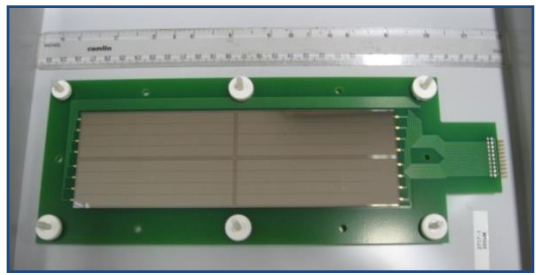

*Fig 5: Segmented silicon detector, assembled at the back of the MWPC*

Fig 5. shows the photograph of the segmented silicon detector made by Micron Semiconductor Ltd, UK as per our design. The window thickness of the silicon detector is 0.3 micron and measured energy resolution (alpha from $^{241}$Am source) was found to be 2.5%.

We have also developed few Parallel Plate Avalanche Counter (PPAC) with active area 5 cm x 3 cm with three electrode geometry. These are used as start detector in a time flight set up. This is particularly required in an accelerator like K130 cyclotron where time resolution of the beam is poor (> 5 ns).

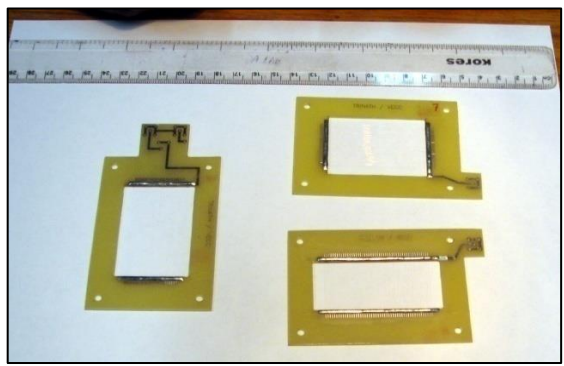

*Fig 6: PCBs of the parallel plate avalanche counters*

The central anode wire plane of the detector consists of 10 μm diameter gold plated tungsten wires soldered 1 mm apart. The cathode wire planes were made of 20 μm diameter gold coated tungsten wire, placed 2 mm apart. The separation between anode and a cathode plane was 3.2 mm. All the wire planes were made of G – 10 quality single sided glass epoxy, copper plated boards (PCB) as shown in Fig 6. The anode wires were soldered on to conducting strips. The cathode wires were similarly soldered to a conducting pad. The two cathode wire planes were shorted outside and connected to a power supply through a charge sensitive pre-amplifier. Stretched polypropylene films of thickness 0.5 micron were used as the entrance windows of the detector. A 1 cm × 1 cm wire mesh of stainless steel wires of diameter 0.4 mm was used as a support to the polypropylene film. Two gas feed-through were connected to the back support frame which is made of stainless steel. In order to keep the distance (3.2 mm) between anode and cathode we had to use spacers (made of G10 board) of thickness 1.6 mm. The detectors are operated at +280 volt (anode) and -180 volt (cathode) with isobutene gas at 3 torr. The detectors provide time resolution better that 500 ps. The detectors are found to be more than 99 % efficient for detection of fission fragments from a $^{252}$Cf source.

At VECC we have also developed an axial field ionization chamber [7]. The detector will be used as a ΔE-detector of a charged particle telescope with silicon strip as E-detector (stop) to study the particle-particle, particle-fragment correla-tion in heavy ion induced reactions. For detection of heavy fragments, employment of this kind of detector is inevitable because of the non-availability of large area silicon detector of sufficiently smaller thickness.

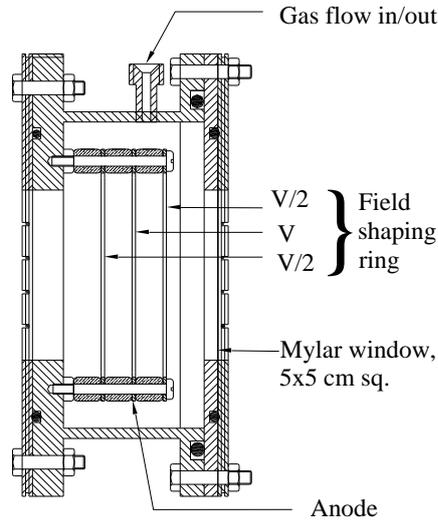

*Fig 7: Vertical cross sectional view of the axial ionisation chamber*

The active area of the detector is 5 cm x 5 cm. The detector consisted of three square shaped field-shaping ring separated by 10 mm. A nichrome wire mesh of 98% transparency was attached with the middle field-shaping ring that acts as an anode. Using a voltage divider resistance chain, a voltage of V is given to anode while V/2 to the other rings. The front and the back of the detector were covered by 2 mm thick mylar foils, glued on stainless steel (SS) frame. The SS plates were grounded. The window foils were supported by thin SS wire of diameter 0.3 mm. The detector is operated with P-10 gas at typical E/p= 1 Volt/cm/torr.

**Development of a Charged Particle Detector Array (CPDA):**
At VECC, we are developing a high-resolution, high-granularity 4-pi array for complete charged particle spectroscopy. This is a unique tool for the study of multi-particle correlation, resonance spectroscopy, complete calorimetry etc. One of the main motivations behind the development of this array is to study the decay of highly excited nuclear system and the search for the density dependence of the symmetry energy part of the nuclear equation of state [8]. The density dependence of the symmetry energy is expected to affect several observables ( e.g; the ratio, between the isotopic yields of a given fragment in two different reactions) which can be measured in heavy-ion collisions. The charged-particle detector array is designed to satisfy the needs of the scientific motivations discussed above, as the array will have excellent angular, energy and isotopic resolutions and large (~ $4\pi$) solid angle coverage with high granularity.

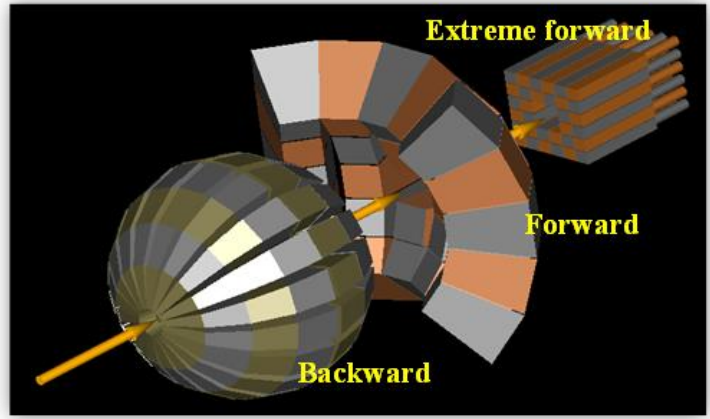

*Fig 8: Design of the Charged Particle Detector Array (CPDA)*

The array consists of three parts: the extreme forward array (angular coverage $\pm 3$ to $\pm 7$ degree), forward array ($\pm 7$ to $\pm 45$ degree) and backward array ($\pm 45$ to $\pm 170$ degree). A schematic view of the charged particle detector array is shown in Fig 8.

The extreme forward part of the array consists of 32 plastic slow (10 cm, BC444 crystal) - fast (100/200 μm, BC408 crystals) phoswich detectors, mostly for the detection of light charged particle, elastic scattering and direct reactions products. These detectors are capable of handling high count rate. They are mounted at 40 cm away from the target that offers angular resolution better than a degree. Typical threshold for these detectors are ~ 2 MeV/nucleon for proton and alpha, ~ 5 MeV/nucleon for $^{16}$O and ~ 8 MeV/nucleon for $^{40}$Ca.

The forward part of the CPDA consists of 24 telescopes, each with silicon strip (50 μm thick) + silicon strip (500 /1000 μm) + 4 CsI(Tl) (6 cm) detectors. The centre of each telescope is mounted 20 cm away from the target position. The silicon detectors (*make Micron semiconductor Ltd, UK*) have dimension of 50 mm x 50 mm, the front one is single sided with 16 strips of 3 mm pitch and the back detector is double sided (each with 16 strips, orthogonal to each other) with same pitch. The measured thickness variation in 50 micron thick silicon detector was found to be better than 3%. Each of the CsI crystal has truncated pyramid shape with front face area 25 mm x 25 mm and back face area 35 x 35 mm. Photo diodes are attached to the back side of each crystal to take out the signals. The measured energy resolutions of the crystal were found to be ~ 5% with $^{241}$Am alpha source. The maximum non-uniformity of the crystal was less than 0.5%.

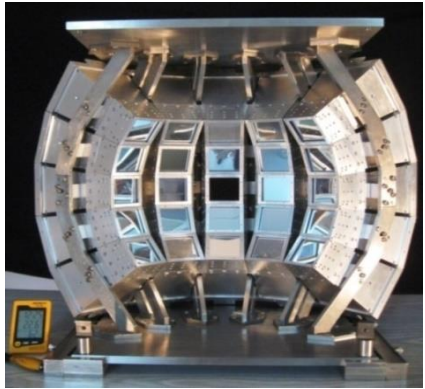

*Fig 9: Photograph of the forward part of the CPDA*

The backward part of the array is to detect only light charged particles (LCP) (Z=1, 2). LCP will be identified using single CsI(Tl) crystal by pulse shape discrimination (PSD) technique. 114 CsI(Tl) detectors will be mounted in 6 rings. Geometry of this part of array is such that front face of the detectors form a part of sphere of radius 15 cm and it does not clash with the other part of the array. The extreme forward and the forward part of the CPDA are ready with analogue electronics (*Mesytec GmbH*). The backward part of the array is currently under development.

The development of CPDA at VECC helped us to carry out several experiments. Part of the CPDA (few telescopes), was used in an experiment to extract the direct component of the decay of the famous Hoyle state [9]. The granularity and required energy resolution allowed to carry out a high-statistics, high-resolution, complete kinematical measurement to estimate the quantitative contributions of various direct 3α decay mechanisms in the decay of the $0_2^+$ resonant excited state of $^{12}$C at an excitation energy of 7.654 MeV, using inelastic scattering of 60-MeV α particles on $^{12}$C. Spectroscopic information about different excited states of $^{26}$Al and $^{26}$Mg populated through the $^{27}$Al(d, t) and $^{27}$Al(d, $^3$He) reactions [10] and the role of α clustering in the binary complex fragment decay of fully energy-relaxed composites $^{24,25}$Mg*, formed in $^{12,13}$C + $^{12}$C reactions, has also been studied using CPDA telescopes [11].

**Development of neutron detectors**:

An array of neutron detectors is being developed at VECC for the spectroscopic measurement of fast neutrons produced in accelerator based experiments. The array will be used for the measurement of energy and angular distribution of neutrons produced in the nuclear collision. This array is planned particularly to study nuclear level density and fusion fission dynamics near and above the Coulomb barrier energies.

The detector array will consist of 50 liquid scintillator (BC501A) based detectors, each of the cell having 5"x5" cylindrical size and is coupled to 5" photo multiplier tube (XP4512B). The centre of each detector will be at a distance of 2 meter from the target position. The target is placed in a 3 mm thin walled spherical scattering chamber of 100 cm diameter. Target chamber also has provision to put large area position sensitive multi-wire proportional counters (MWPC) to detect fission fragments in coincidence with neutrons.

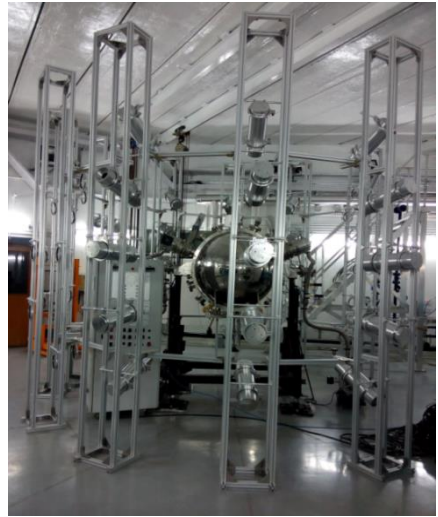

*Fig 10: Photograph of the detector array around the thin wall scattering chamber at K500 Super Conducting beam hall at VECC*

For such measurements, the TOF neutron detector array is required to have good pulse shape discrimination, intrinsic time resolution and efficiency over the range of neutron energies from about 1 MeV to few tens of MeV. This developmental activity was guided by several R&D activities which includes development of detectors of several size (7 inch, 5 inch , 3 inch and 1.5 inch length with 5 inch diameter) and exploration of their detection properties (figure of merit, time resolution and efficiency) to select the optimum detector that suits our requirement . With increase in detector volume, figure of merit (1.26 for 1.5 inch to 1.07 for 7 inch) and time resolution (1.44 ns for 1.5 inch to 1.56 ns for 7 inch) worsen as expected, which is due to higher dimension associated with larger light loss and larger time spread of the arrival of photon at PMT [12]. However, the efficiency of the detector increases with the increase in detector volume. Considering all aspects, for our array, 5 inch x 5 inch detector was chosen that offers time resolution (FWHM) ~ 1.53 ns, figure of merit (with threshold 350 keVee) 1.08 and efficiency (at 2 MeV, threshold 100 keVee) ~ 64 %. A photograph of the detector array around the thin wall scattering chamber at K500 Super Conducting beam hall at VECC is shown in Fig 10.

The detectors can easily be transported to different experimental area. Few such detectors are routinely used in in-beam experiments performed at K130 Cyclotron in VECC. Angular momentum dependence of nuclear level density [13] out of collectivity in nuclear level density [14], shell effect in nuclear level density in $^{208}$Pb region [15] were extensively studied using these detectors.

**Summary:**
In summary, several state of the art detectors are being developed at VECC. The detectors are being effectively utilised to study fundamental Nuclear Physics such as the study of multi-particle correlations, fragment emission mechanism, study of nuclear level density and fusion fission dynamics.

**Acknowledgements:** The major part of this work was carried out under the Super Conducting Utilisation Project. The author is thankful to S. Bhattacharya, C. Bhattacharya, K. Banerjee, S. Kundu, T.K. Rana, J.K. Meena, A. Chaudhuri, G. Mukherjee, P. Roy, R. Pandey, S. Manna, A. Sen, J. K. Sahoo, A. Saha, R. Saha Mondal, Trinath Banik, Arnab Ghosh, who have contributed in this work. Sincere thanks to Prof. S. Bhattacharya for his suggestions to prepare of the manuscripts.

**References:**
[1] J.C. Bose, Collected Physical Papers. New York, N.Y.: Longmans, Green and Co., (1927)
[2] C. B. Das *et al*., Phys. Rep. 406, 1 (2005)
[3] J. Mabiala *et al*., Journal of Physics: Conference Series, 420(1) (2013).
[4] T.K. Ghosh, Pramana – J. Phys. 85, 291 (2015); A. Chaudhuri *et al*., Phys. Rev. C 92, 041601 (2015) (R)
[5] Charpak and Sauli , Nucl. Instrum. Methods 162, 405 (1979)
[6] T.K. Ghosh *et al*., Nucl. Instrum. and Methods in Phys Res A 540, 285 (2005)
[7] S.K. Bandopadhyay *et al*., Nucl. Instrum. and Methods in Phys Res A 278, 467 (1989)
[8] P. Danielewicz and J. Lee, Int. J. Mod. Phys. E 18, 892 (2009).
[9] T.K. Rana *et al*., Phys. Rev. C 88, 021601 (2013) (R)
[10] V. Srivastava *et al*., Phys. Rev. C 93, 044601 (2016), Phys. Rev. C 91, 054611 (2015)
[11] S. Manna *et al*., Phys. Rev. C 94, 051601 (2016) (R)
[12] K. Banerjee *et al*., Nucl. Instrum. and Methods in Phys Res A 608, 440 (2009)
[13] K. Banerjee *et al*., Phys. Rev. C 85, 064310 (2012)
[14] Pratap Roy *et al*., Phys. Rev. C 88, 031601 (2013) (R)
[15] Pratap Roy *et al*., Phys. Rev. C 94, 064607 (2016)